\documentclass[prl,aps,twocolumn]{revtex4}

\usepackage{amsmath}
\usepackage{bm}
\usepackage{graphicx}

\begin{document}

\title{Breaking of Chiral Symmetry and Spontaneous Rotation in a Spinor
Bose-Einstein Condensate}

\author{Hiroki Saito$^1$}
\author{Yuki Kawaguchi$^1$}
\author{Masahito Ueda$^{1,2}$}
\affiliation{$^1$Department of Physics, Tokyo Institute of Technology,
Tokyo 152-8551, Japan \\
$^2$ERATO, Japan Science and Technology Corporation (JST), Saitama
332-0012, Japan
}

\date{\today}

\begin{abstract}
We show that a spin-1 Bose-Einstein condensate with ferromagnetic
interactions spontaneously generates a topological spin texture, in which
the $m = \pm 1$ components of the magnetic sublevels form vortices with
opposite circulations.
This phenomenon originates from an interplay between ferromagnetic
interactions and spin conservation.
\end{abstract}


\maketitle

The Mermin-Ho texture~\cite{Mermin} in superfluid $^3{\rm He}$ describes
an interesting thermodynamic equilibrium state, in which a circulation
remains nonvanishing in a cylindrically symmetric vessel, despite the
vessel being at rest.
This phenomenon is due to the fact that the boundary condition at the
surface of the vessel imposes a topological structure on the $\bm{l}$
vector.
In this Letter, we show that the spontaneous formation of such a
topological spin texture also occurs in a Bose-Einstein condensate (BEC)
of atomic gases with spin degrees of freedom~\cite{Kurn,Ohmi,Ho}.

In contrast to the Mermin-Ho texture in $^3{\rm He}$, the physical origin
of the spin texture formation proposed in this Letter is the interplay
between ferromagnetic interactions and spin conservation.
Consider a spin-1 BEC with ferromagnetic interactions in an $m = 0$
magnetic sublevel.
The spin-exchange collisions between the atoms transfer the $m = 0$
state into $m = \pm 1$ states, as $0 + 0 \rightarrow 1 + (-1)$.
As a consequence, magnetization in the $x$-$y$ plane may arise due to
the ferromagnetic nature of the interaction.
However, uniform magnetization of the entire system is prohibited because
of spin conservation, which results in various spin
textures~\cite{Saito,Zhang,Petit}, such as staggered domain structures.
In this Letter, we show that the topological spin texture is spontaneously
generated as a result of the spin exchange dynamics under spin
conservation.

In this topological spin texture, each of the $m = \pm 1$ components
contains a vortex, whose directions are opposite.
Therefore, from the symmetry of the Hamiltonian, there are two degenerate
textures: one has $+$ and $-$ vorticies in the $m = 1$ and $-1$
components, respectively, and the other has $-$ and $+$ vortices.
We show that the symmetry between these two textures is spontaneously
broken in the course of the dynamics even when the initial state has
chiral symmetry.

Topological spin structures in a spinor BEC of an atomic gas have been
realized by the MIT group~\cite{Leanhardt}, who obtained topological
phases adiabatically imprinted on the spin components by a quadrupole
magnetic field, producing coreless vortices.
The stability of such topological spin structures has been studied by
several authors~\cite{Alkhawaja,Mizushima,Savage}.
Recently, it was predicted~\cite{Santos,Kawaguchi} that the dipolar
interaction creates a coreless vortex state through a mechanism similar to
the Einstein-de Haas effect.

We consider a system of spin-1 Bose atoms with mass $M$ confined in a
potential $V$.
The Hamiltonian of the system is given by~\cite{Ohmi,Ho}
\begin{eqnarray} \label{H}
\hat H & = & \int d\bm{r} \Biggl[ \sum_m \hat\psi_m^\dagger H_0 \hat\psi_m
+ \sum_{m_{1,2,3,4}} \hat\psi_{m_1}^\dagger \hat\psi_{m_2}^\dagger
\nonumber \\
& & \times \left( \frac{g_0}{2} \delta_{m_1 m_4} \delta_{m_2 m_3} 
+ \frac{g_1}{2} \bm{F}_{m_1 m_4} \cdot \bm{F}_{m_2 m_3} \right)
\hat\psi_{m_3} \hat\psi_{m_4} \Biggr], \nonumber \\
\end{eqnarray}
where $\hat\psi_m$ is the field operator for an atom in the magnetic
sublevel $m = 0$, $\pm 1$, $H_0 = -\hbar^2 \nabla^2 / (2 M) + V$, and
$\bm{F}$ is the spin-1 matrix.
The spin-independent and spin-dependent interactions are characterized by
$g_0 = 4 \pi \hbar^2 (a_0 + 2 a_2) / (3 M)$ and $g_1 = 4 \pi \hbar^2 (a_2
- a_0) / (3 M)$, respectively, where $a_S$ is the $s$-wave scattering
length for the scattering channel with total spin $S$.
In the case of spin-1 $^{87}{\rm Rb}$, $g_1$ is negative and the ground
state is ferromagnetic~\cite{Ho,Klausen}.

When the potential $V$ is axisymmetric with respect to the $z$ axis, the
Hamiltonian~(\ref{H}) is invariant under spatial reflection with respect
to an arbitrary plane containing the $z$ axis, e.g., $(x, y) \rightarrow
(-x, y)$.
This transformation changes a clockwise vortex $\propto e^{-i \phi}$ into
a counterclockwise vortex $\propto e^{i \phi}$ with azimuthal angle
$\phi$.
Hence, if only one of them is spontaneously realized, we call
it chiral symmetry breaking.
Also, from the symmetry of the Hamiltonian, the $x$, $y$, and $z$
components of the total spin and the $z$ component of the orbital angular
momentum are conserved independently.

We consider the case in which the initial state is in the $m = 0$
mean-field ground state satisfying
\begin{equation} \label{psi0}
\left(H_0 + g_0 |\Psi_0|^2 \right) \Psi_0 = \mu_0 \Psi_0.
\end{equation}
If the $m = \pm 1$ components are exactly zero, $\Psi_0$ is a stationary
state of the multicomponent Gross-Pitaevskii (GP) equations,
\begin{subequations} \label{GP}
\begin{eqnarray}
i \hbar \frac{\partial \psi_0}{\partial t} & = & \left(H_0 + g_0 n \right)
\psi_0 + \frac{g_1}{\sqrt{2}} \left( F_+ \psi_1 + F_- \psi_{-1} \right),
\label{GP1} \\
i \hbar \frac{\partial \psi_{\pm 1}}{\partial t} & = & \left( H_0 + g_0 n
\right) \psi_{\pm 1} + g_1 \left( \frac{1}{\sqrt{2}} F_{\mp} \psi_0 \pm
F_z \psi_{\pm 1} \right), \nonumber \\
\label{GP2}
\end{eqnarray}
\end{subequations}
where $\psi_m$ is the macroscopic wave function, $n = \sum_m |\psi_m|^2$,
$F_z = |\psi_1|^2 - |\psi_{-1}|^2$, and $F_+ = F_-^* = \sqrt{2} (\psi_1^*
\psi_0 + \psi_0^* \psi_{-1})$.
The stability against excitation in the $m = \pm 1$ components is
analyzed by the Bogoliubov-de Gennes equations, given by
\begin{subequations} \label{Bogo}
\begin{eqnarray}
\left[ H_0 - \mu_0 + (g_0 + g_1) |\Psi_0|^2 \right] u_{\pm 1}^{(\ell)} +
g_1 \Psi_0^2 v_{\mp 1}^{(\ell) *} & = & \varepsilon^{(\ell)} u_{\pm
1}^{(\ell)}, 
\nonumber \\
\\
\left[ H_0 - \mu_0 + (g_0 + g_1) |\Psi_0|^2 \right] v_{\mp 1}^{(\ell)*} +
g_1 \Psi_0^{*2} u_{\pm 1}^{(\ell)} & = & -\varepsilon^{(\ell)} v_{\mp
1}^{(\ell) *}, \nonumber \\
\end{eqnarray}
\end{subequations}
where $u_m^{(\ell)}$ and $v_m^{(\ell)}$ are the eigenfunctions for a
Bogoliubov mode with eigenenergy $\varepsilon^{(\ell)}$.
From the axisymmetry of the system, the Bogoliubov modes can be classified
according to the angular momentum $\ell$, for which $u_m^{(\ell)} \propto
e^{i \ell \phi}$ and $v_m^{(\ell)} \propto e^{-i \ell \phi}$.
We find that $u_1^{(\ell)}$ couples with $v_{-1}^{(\ell)*}$ in
Eq.~(\ref{Bogo}), and hence the excitation of the $m = 1$ component with
vorticity $\ell$ is accompanied by the $m = -1$ component with vorticity
$-\ell$, as a consequence of the orbital angular momentum conservation.
If all the eigenenergies are real, the state $\Psi_0$ is dynamically
stable.
If there exist complex eigenenergies, the corresponding modes grow
exponentially and the state $\Psi_0$ is dynamically unstable.

For simplicity, we restrict ourselves to two-dimensional (2D) space.
This situation can be realized by a tight pancake-shaped potential $V = M
\omega^2 (x^2 + y^2 + \lambda^2 z^2) / 2$ with $\lambda \gg 1$, where the
axial confinement energy is so large that the dynamics in the $z$
direction are frozen.
In this case, the interaction strengths can be characterized by the
dimensionless parameters~\cite{Castin} $g_j^{\rm 2D} = g_j [\lambda / (2
\pi)]^{1/2} N / (a_{\rm ho}^3 \hbar \omega)$, where $N$ is the number of
atoms, $a_{\rm ho} = [\hbar / (M \omega)]^{1/2}$, and $j = 1$, $2$.

\begin{figure}[tb]
\includegraphics[width=8.4cm]{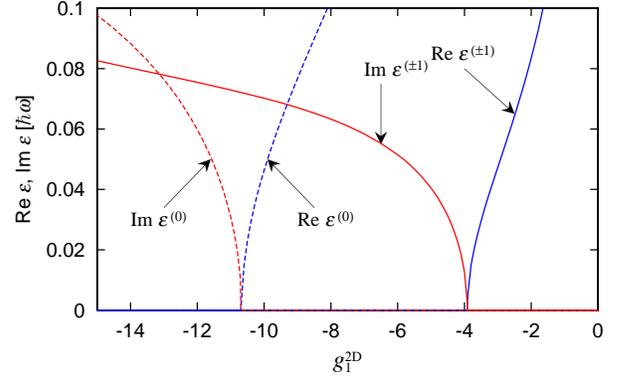}
\caption{
(Color) Real and imaginary parts of the lowest Bogoliubov energies
$\varepsilon^{(\ell)}$ for $\ell = 0$, $\pm 1$, where the label $\ell$
indicates that the $m = \pm 1$ components of the eigenfunction are
proportional to $e^{\pm i \ell \phi}$.
The two energies $\varepsilon^{(\pm 1)}$ are degenerate due to the
axisymmetry of the system.
We have taken the parameters of spin-1 $^{87}{\rm Rb}$ atoms, where the
spin-independent interaction strength $g_0^{\rm 2D}$ is related to the
spin-dependent strength $g_1^{\rm 2D}$ by $g_0^{\rm 2D} = -216.1 g_1^{\rm
2D}$.
}
\label{f:spectrum}
\end{figure}

We numerically solve Eq.~(\ref{psi0}) by the imaginary-time propagation
method and diagonalize Eq.~(\ref{Bogo}) to obtain the Bogoliubov spectrum
for the state $\Psi_0$.
Figure~\ref{f:spectrum} shows the lowest Bogoliubov energies for $\ell =
0$ and $\pm 1$ as a function of $g_1^{\rm 2D}$, where $g_0^{\rm 2D}$ is
determined by $g_0^{\rm 2D} / g_1^{\rm 2D} = g_0 / g_1 \simeq -216.1$
which is the ratio for spin-1 $^{87}{\rm Rb}$~\cite{Kempen}.
In the parameter regime shown in Fig.~\ref{f:spectrum}, the three modes
exhibit complex eigenenergies.
A crucial observation is that there is a region ($-3.9 \gtrsim g_1^{\rm
2D} \gtrsim -10.7$) in which only $\varepsilon^{(\pm 1)}$ are imaginary.
This indicates that these two modes only are dynamically unstable in this
region, where one mode has vortices $\propto e^{\pm i \phi}$ and the
other mode has vortices $\propto e^{\mp i \phi}$ in the $m = \pm 1$
components.
These two modes are degenerate because of the chiral symmetry of the
system.
In this region, we expect that the $m = \pm 1$ components start to rotate,
despite there being no external rotating drive is applied to the system.

\begin{figure}[tb]
\includegraphics[width=8.4cm]{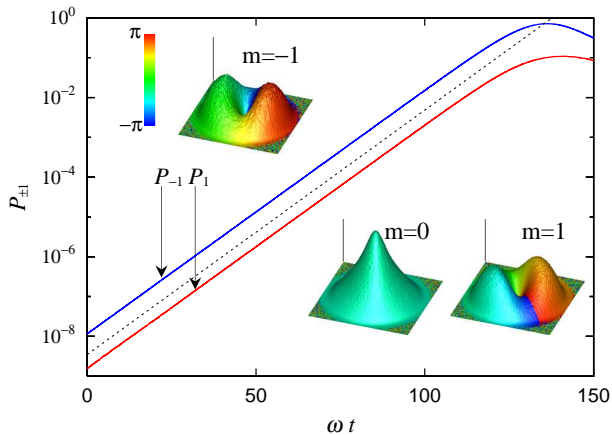}
\caption{
(Color) Degree of Bogoliubov excitation $P_{\pm 1}$ in Eq.~(\ref{P}).
The interaction strengths are $g_0^{\rm 2D} = 2200$ and $g_1^{\rm 2D} =
-10.18$.
The initial state is the ground state $\Psi_0$ of Eq.~(\ref{psi0}) for the
$m = 0$ component plus a small amount of random noise in the $m = -1$
component.
The dashed line is proportional to $e^{0.141 \omega t}$.
The insets show the density-phase profiles at $\omega t = 130$, where
the size of the frame is $16 \times 16$ in units of $a_{\rm ho} = [\hbar /
(M \omega)]^{1/2}$.
The heights of the vertical lines in the insets indicate $|\psi_m|^2
a_{\rm ho}^2 / N$, which is $0.001$ for $m = \pm 1$ and $0.0025$ for
$m = 0$.
}
\label{f:noise}
\end{figure}

In order to confirm the spontaneous rotation phenomenon predicted above,
we numerically solve the GP equation~(\ref{GP}) in 2D using the
Crank-Nicholson scheme.
The interaction strengths are taken to be $g_0^{\rm 2D} = 2200$ and
$g_1^{\rm 2D} = -10.18$, with the ratio $g_0^{\rm 2D} / g_1^{\rm 2D}$
again chosen to be that of spin-1 $^{87}{\rm Rb}$.
For this set of interaction parameters, ${\rm Im} \varepsilon^{(\pm 1)} /
(\hbar \omega) = 0.0707$ and all the other Bogoliubov energies are real.
The initial state is the ground state of Eq.~(\ref{psi0}) for the $m = 0$
component plus a small amount of random noise in the $m = -1$ component.
To extract the Bogoliubov excitations from $\psi_m(t)$, we
define~\cite{Kawaguchi04}
\begin{equation} \label{P}
P_{\pm 1} \equiv \left| \int d\bm{r} \left[ e^{i \mu_0 t / \hbar}
u_1^{(\mp 1)} \psi_1(t) - e^{-i \mu_0 t / \hbar} v_{-1}^{(\mp 1)}
\psi_{-1}^*(t) \right] \right|^2,
\end{equation}
which represents the degree of excitation in the modes $u_1^{(\pm 1)}
\propto v_{-1}^{(\pm 1)*} \propto e^{\pm i \phi}$.
Figure \ref{f:noise} shows the time evolution of $P_{\pm 1}$ and
the density-phase profile of each component at $\omega t = 130$.
We find that $P_{\pm 1}$ increase according to $\exp(2 {\rm Im}
\varepsilon^{(\pm 1)} t / \hbar)$, with their initial ratio $P_{-1} / P_1
\simeq 7.5$ kept constant, resulting in an exponential growth in the
angular momenta of the $m = \pm 1$ components.
Thus, the $m = \pm 1$ components spontaneously rotate if the initial
noise has angular-momentum fluctuations.

\begin{figure}[tb]
\includegraphics[width=8.4cm]{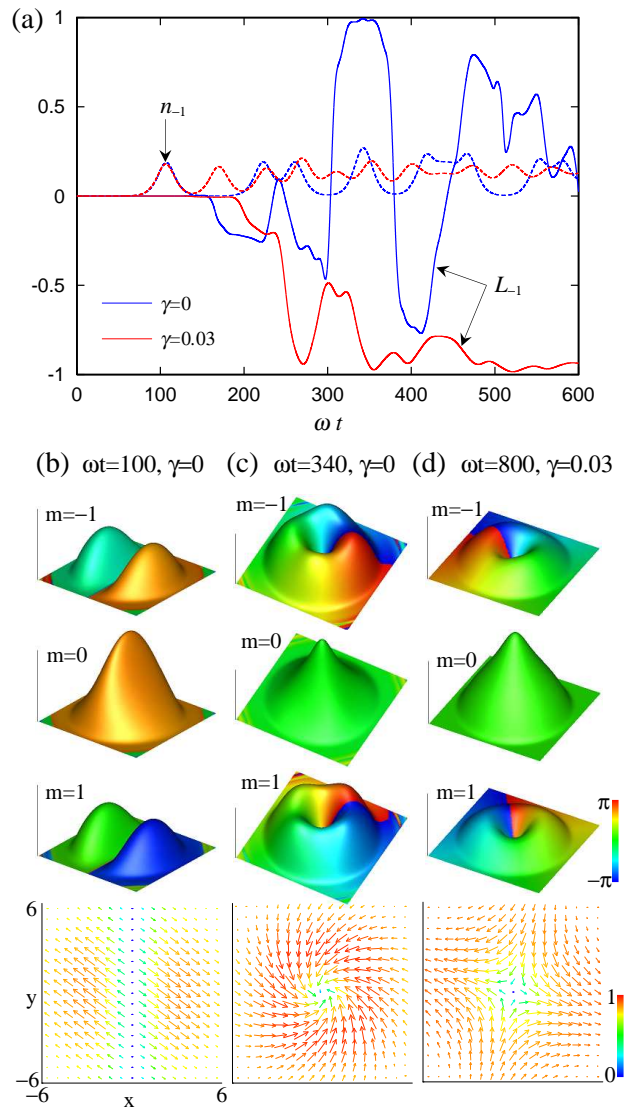}
\caption{
(Color) (a) Time evolution of the fraction $n_{-1}$ (dashed curves) and
the orbital angular momentum per particle $L_{-1}$ (solid curves) in the
$m = -1$ component with ($\gamma = 0.03$, red) and without ($\gamma = 0$,
blue) dissipation.
The interaction strengths are the same as in Fig.~\ref{f:noise}.
The initial state is given by $\psi_0 = \Psi_0$, $\psi_{-1} = 10^{-4} r
(e^{i \phi} + 1.0001 e^{-i \phi}) \Psi_0$, and $\psi_1 = 0$, where
$\Psi_0$ is the ground state solution of Eq.~(\ref{psi0}).
(b)-(d) Snapshots of the density-phase profiles and spin textures.
The heights of the vertical lines in the density-phase profiles show
$|\psi_m|^2 a_{\rm ho} / N$, which is $0.001$ for $m = \pm 1$ and $0.0025$
for $m = 0$.
The length of the vector in the bottom panels is proportional to $(F_x^2 +
F_y^2)^{1/2}$ and the color represents $(F_x^2 + F_y^2)^{1/2} /
(|\psi_{-1}|^2 + |\psi_0|^2 + |\psi_1|^2)$.
The size of the frame is $16 \times 16$ for the density-phase profiles and
$12 \times 12$ for the spin textures in units of $a_{\rm ho}$.
}
\label{f:longtime}
\end{figure}

Suppose that we can prepare an initial state of the system in which the
chiral symmetry is preserved to great accuracy, say, $P_{-1} / P_1 =
1.0002$.
Then, the result in Fig.~\ref{f:noise} indicates that vortices will not
be created in the time scale in which the linear stability analysis is
applicable.
For a longer time scale, however, the chiral symmetry is spontaneously
broken due to the nonlinear effect.
Figure~\ref{f:longtime} (a) shows the time evolution of the fraction
$n_{-1} = \int d\bm{r} |\psi_{-1}|^2 / N$ and the orbital angular momentum
per particle $L_{-1} = -i \int d \bm{r} \psi_{-1}^* \partial_\phi
\psi_{-1} / (N n_{-1})$ for the $m = -1$ component.
The $m = 1$ component is given by $n_1 \simeq n_{-1}$ and $L_1 \simeq
-L_{-1}$ from spin and orbital angular momentum conservation.
The initial value for the $m = -1$ component is taken to be $\psi_{-1} =
10^{-4} r (e^{i \phi} + 1.0001 e^{-i \phi}) \Psi_0$, which gives $P_{-1} /
P_1 \simeq 1.0002$.
As long as this ratio is kept constant, the formation of the vortex states
shown in the insets of Fig.~\ref{f:noise} is not expected.
In fact, as shown in Fig.~\ref{f:longtime} (b), no vortex is created
around the first peak of $n_{-1}$ at $\omega t \simeq 100$.
However, at $\omega t \simeq 160$, $L_{-1}$ starts to deviate from 0 (blue
solid curve) and the chiral symmetry is dynamically broken.
Consequently, the vortex states emerge in the $m = \pm 1$ components at
$\omega t \simeq 340$ as shown in Fig.~\ref{f:longtime} (c).

The instability in the state with chiral symmetry [Fig.~\ref{f:longtime}
(b)] against forming the rotating state [Fig.~\ref{f:longtime} (c)]
implies that the energy of the latter is lower than that of the former.
In order to confirm this, we take the energy dissipation into account by
replacing $i$ on the left-hand side of Eq.~(\ref{GP}) with $i -
\gamma$~\cite{Tsubota}.
The time evolution with $\gamma = 0.03$~\cite{Choi} is shown by the red
curves in Fig.~\ref{f:longtime} (a) and clearly indicates that the
energy of the rotating state [Fig.~\ref{f:longtime} (d)] is lower than
that of the state having chiral symmetry [Fig.~\ref{f:longtime} (b)].
For $\gamma = 0$, $L_{-1}$ oscillates with a large amplitude due to the
excess energy released from the initial state, while for $\gamma = 0.03$,
the sign of the angular momentum is unchanged.

The bottom panels in Figs.~\ref{f:longtime} (b)-(d) show the spin vector
distributions.
In Fig.~\ref{f:longtime} (b), the magnetic domains in the opposite spin
directions are separated by a domain wall at $x = 0$.
On the other hand, topological spin structures are formed in
Figs.~\ref{f:longtime} (c) and (d).
The underlying physics of the spin structure formation is the interplay
between the ferromagnetic interaction and spin conservation.
The growth in the spin vectors must be accompanied by 
spatial spin structure formation to conserve the total spin angular
momentum.
It should be noted that the area in which the length of the spin vector is
long is larger in Figs.~\ref{f:longtime} (c) and (d) than in
Fig.~\ref{f:longtime} (b), since the spin vectors must vanish at the
domain wall in the latter.
This is why the energy of the state in Fig.~\ref{f:longtime} (d) is lower
than that of Fig.~\ref{f:longtime} (b).
That is, the formation of the topological spin structure increases the
(negative) ferromagnetic energy of the system more than does the formation
of the domain structure.

\begin{figure}[tb]
\includegraphics[width=8.4cm]{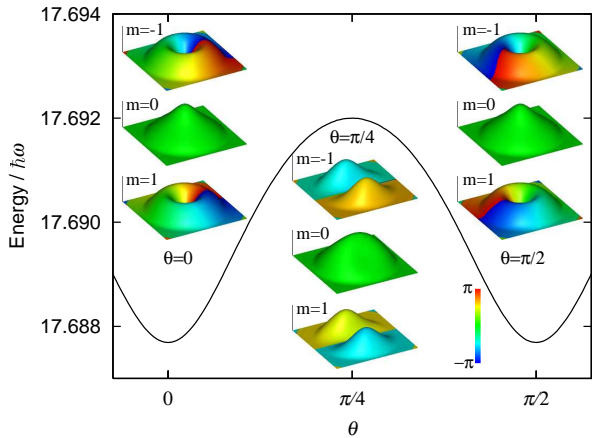}
\caption{
(Color) Energy and density-phase profiles obtained by the variational
	method with Eq.~(\ref{var}).
The interaction strengths are the same as in Fig.~\ref{f:noise}.
The size of the frame of the insets is $16 \times 16$ in units of $a_{\rm
ho}$ and the heights of the vertical lines show $|\psi_m|^2 a_{\rm ho}^2
/ N$, which is $0.0005$ for $m = \pm 1$ and $0.0025$ for $m = 0$.
}
\label{f:energy}
\end{figure}

The above energy argument concerning the spin domain and topological
structures can be reinforced by applying the variational method.
We assume the variational wave function to be
\begin{equation} \label{var}
\left( \begin{array}{c} 0 \\ \tilde\Psi_0 \\ 0 \end{array} \right) + c
\cos\theta \left( \begin{array}{c} u_1^{(1)} \\ 0 \\ e^{i \chi}
v_{-1}^{(1)} \end{array} \right) + c \sin\theta \left( \begin{array}{c}
u_1^{(-1)} \\ 0 \\ e^{i \chi'} v_{-1}^{(-1)} \end{array} \right),
\end{equation}
where $\tilde\Psi_0 = (|\Psi_0|^2 - |\psi_1|^2 - |\psi_{-1}|^2)^{1/2}$ so
that the total density $n$ is kept to be $|\Psi_0|^2$ irrespective of the
values of the variational parameters, reflecting the fact that the
spin-exchange process hardly changes the total density because $g_0 \gg
|g_1|$ for spin-1 $^{87}{\rm Rb}$ atoms.
We minimize the energy of the system calculated from Eq.~(\ref{var}) with
respect to $c$, $\chi$, and $\chi'$ for a given $\theta$, as shown in
Fig.~\ref{f:energy}.
The state at $\theta = \pi / 4$, which is similar to the state shown in
Fig.~\ref{f:longtime} (b), has a maximum energy, while the topological spin
states at $\theta = 0$ and $\pi / 2$ have minimum energy, in agreement
with the above discussion.

The results presented above can be realized using current experimental
setups.
For example, when the radial trapping frequency is $\omega = 100 \times 2
\pi$ Hz and its ratio to the axial trapping frequency is $\omega /
\omega_z = 0.01$, the interaction parameters for
Figs.~\ref{f:noise}-\ref{f:energy} correspond to $N \simeq 8880$ spin-1
$^{87}{\rm Rb}$ atoms.
The time scale for the appearance of the topological spin structure (e.g.,
$\omega t \sim 300$ for the initial condition in Fig.~\ref{f:longtime})
is $\sim 0.5$ s.
If the ratio $|g_1 / g_0|$ can be increased with a decrease in $g_0$ by
Feshbach resonance or by using other atomic species, we can reduce the
time scale, e.g., to about $1 / 10$ for $|g_1 / g_0| = 1$.

In the presence of an external magnetic field $B$, the linear and
quadratic Zeeman terms enter the Hamiltonian (\ref{H}).
Since the total spin is conserved, the linear Zeeman term only rotates the
spin at the Larmor frequency and does not affect the dynamics.
When the magnetic field is applied in the $z$ direction, the quadratic
Zeeman effect raises the energy of the $m = \pm 1$ components compared
with the $m = 0$ component.
If the quadratic Zeeman energy exceeds the ferromagnetic energy, the $m =
0$ state becomes the ground state and no excitation to the $m = \pm 1$
components occurs, which is the case for $B \gtrsim 400$ mG for the
parameters $\omega = 100 \times 2 \pi$ Hz, $\omega / \omega_z = 0.01$, and
$N \simeq 8880$.
We have numerically confirmed that the dynamics in Figs.~\ref{f:noise} and
\ref{f:longtime} are qualitatively unchanged for a magnetic field of
$\simeq 100$ mG.

In conclusion, we have proposed a novel mechanism of spontaneous formation
of a topological spin structure in the spin-1 BEC prepared in the $m = 0$
state.
The $m = \pm 1$ components increase exponentially from initial random
seeds due to dynamical instabilities and form singly-quantized vortex
states (Fig.~\ref{f:noise}).
Even if the clockwise and counterclockwise rotation components are
assumed to be equal in an initial seed, one of them eventually becomes
dominant (Fig.~\ref{f:longtime}).
This chiral symmetry breaking is attributed to the fact that the
topological spin structure is energetically the most favorable due to the
ferromagnetic interaction.
This spontaneous spin structure formation is essentially caused by the
spin exchange dynamics under the constraint of spin conservation.
We expect that many more interesting spin textures may also be
spontaneously generated in isolated spinor BECs.

This work was supported by Grant-in-Aids for Scientific Research (Grant
No.\ 17740263, No.\ 17071005, and No.\ 15340129) and by a 21st Century COE
program at Tokyo Tech ``Nanometer-Scale Quantum Physics,'' from the
Ministry of Education, Culture, Sports, Science and Technology of Japan.
YK acknowledges support by a Fellowship Program of the Japan Society
for Promotion of Science (Project No. 16-0648).
MU acknowledges support by a CREST program of the JST.

\end{document}